# Graphene under hydrostatic pressure

John E. Proctor[1]*, Eugene Gregoryanz[1], Konstantin S. Novoselov[2], Mustafa Lotya[3], Jonathan N. Coleman[3], and Matthew P. Halsall[4]

[1]School of Physics and Centre for Science at Extreme Conditions, University of Edinburgh, Edinburgh EH9 3JZ, UK
[2]School of Physics and Astronomy, University of Manchester, Oxford Road, Manchester M13 9PL, UK
[3]School of Physics, Trinity College Dublin, Dublin 2, Ireland
[4]School of Electrical and Electronic Engineering, Photon Science Institute, University of Manchester, Oxford Road, Manchester M13 9PL, UK

* Corresponding author:  jproctor@staffmail.ed.ac.uk

## Abstract

*In-situ* high pressure Raman spectroscopy is used to study monolayer, bilayer and few-layer graphene samples supported on silicon in a diamond anvil cell to 3.5 GPa. The results show that monolayer graphene adheres to the silicon substrate under compressive stress. A clear trend in this behaviour as a function of graphene sample thickness is observed. We also study unsupported graphene samples in a diamond anvil cell to 8 GPa, and show that the properties of graphene under compression are intrinsically similar to graphite. Our results demonstrate the differing effects of uniaxial and biaxial strain on the electronic bandstructure.

## Article

The discovery of graphene in 2004 [1] has led to many advances in solid state physics. Research into this new material is fuelled by interest in fundamental physics as the quantum Hall effect has been observed in graphene at room temperature [2] and electrons within graphene behave as massless dirac fermions, mimicking relativistic particles [3]. Graphene has been suggested as a candidate for a wide variety of applications in electronics (due to its ballistic transport at room temperature) and composite materials [2]. It is the first experimental realisation of a truly 2-dimensional material.

To date there have been no studies published on graphene at high pressure. This is surprising in view of the huge interest in the mechanical properties of graphene [4-9] motivated particularly by its possible applications in nanoelectronics [4, 5]. Strain monitoring is of critical importance [10, 11] in this field. It should be of particular relevance in the case of graphene due to the predicted dependence of electronic bandgap on strain [12], and also due to the fact that some of the materials related to graphene are intrinsically stressed due to the presence of the substrate, for example graphene grown epitaxially on SiC [13]. The possibility of using graphene as an ultrasensitive strain sensor has also been suggested [9]. Study of graphene at high pressure therefore has the potential to develop into an important component in the characterization and understanding of this remarkable new material, as has been the case with carbon nanotubes.

Experiments under hydrostatic pressure have been extensively employed to probe basic characteristics of carbon nanotubes such as compressibility [14], the nature of the nanotube bundle [15] and electronic bandstructure [16]. Further experiments have been motivated by desire to understand the characteristics of composite materials containing nanotubes [17]. The possibilities for using fullerenes and nanotubes for the synthesis of superhard materials at high pressure and temperature, and for hydrogen storage, are also being explored [18, 19].

In this paper we present the first study of graphene at high pressure. Samples of monolayer, bilayer and few-layer graphene supported on silicon are studied, along with unsupported graphene samples. We perform simple calculations to compare our results to those of the recent experiments on graphene under uniaxial strain [8] and the hydrostatic pressure experiments on graphite [20, 21].

The high pressure Raman measurements presented in this paper were performed in gasketed symmetric diamond anvil cells (DACs) and recorded using a micro-Raman spectrometer, at room temperature. Scattered light from the sample was collected in the backscattering geometry and the 514.5 nm radiation of an $Ar^+$ laser was used for excitation throughout. The laser power reaching the DAC did not exceed 20 mW. A lower power level (3 mW reaching the sample) was used for spectra taken in air due to the risk of heating and oxidising the samples. Pressure was recorded using the ruby fluorescence method and nitrogen was used as the pressure-transmitting medium except where otherwise indicated, ensuring quasi-hydrostatic conditions.

Supported graphene samples were prepared using the mechanical exfoliation technique [1], on 100 µm thick silicon wafers coated with a 300 nm thick $SiO_2$ layer, and unsupported graphene samples were prepared using the liquid-phase exfoliation technique [22]. The unsupported graphene samples are a mixture of monolayer, bilayer and few-layer graphene and also contain a small amount of nanographite. Details of the sample preparation are given in the supplementary information.

At low pressures (below ≈0.5 GPa) the D* Raman peak from the graphene samples overlapped partially with the 2nd order Raman peak from the diamond anvils of the high pressure cell, so to obtain the D* Raman peak from the samples we subtracted the background Raman signal from the diamond. See example spectra in the supplementary information.

To complement our experimental data, we perform simple calculations to predict the pressure dependence of the in-plane Raman modes of monolayer graphene. For 3-dimensional, isotropic materials the pressure-induced shift of each Raman mode is related to the compression of the material by the mode Grüneisen parameter γ [23]

$$\omega(P)/\omega_0 = [V(P)/V_0]^{-\gamma} \quad (1)$$

However, in graphite the in-plane compressibility (a axis) is an order of magnitude lower than the out-of-plane compressibility (c axis) [20] so it is preferable to define

Grüneisen parameters for the in-plane and out-of-plane vibrational modes separately using the linear compressibility along the a and c axes respectively. For the in-plane $E_{2g}$ mode at 1580 cm$^{-1}$ under hydrostatic pressure or in-plane biaxial compression one should write

$$\omega(P)/\omega_0 = [a(P)/a_0]^{-2\gamma_{E_{2g}}} \qquad (2)$$

For uniaxial compression one should write

$$\omega(P)/\omega_0 = [a(P)/a_0]^{-\gamma_{E_{2g}}} \qquad (3)$$

This follows the approach of Refs. 20 and 24 and is the definition used in the investigations of graphene under uniaxial strain in Ref. 8 where the Grüneisen parameters are calculated for the $E_{2g}$ (G peak, 1583 cm$^{-1}$) and $A_{1g}$ (D* peak, 2680 cm$^{-1}$) in-plane Raman modes of graphene. Assuming that the in-plane compressibility of graphene is the same as for graphite (given in Ref. 20) we can use equation (2) with the Grüneisen parameters calculated for graphene in Ref. 8 to predict the pressure-induced shifts of the in-plane Raman modes of unsupported graphene.

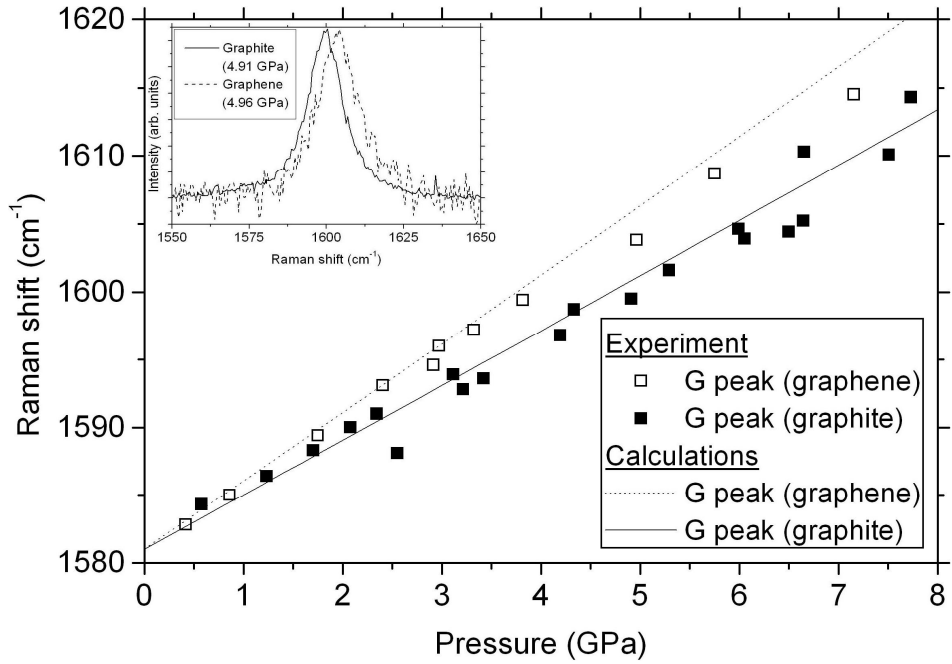

Figure 1. The evolution of the Raman G peak is shown to 8 GPa for unsupported graphene (open squares) and for graphite (filled squares). Our calculations are shown for graphene (dotted line) and graphite (continuous line) using the Grüneisen parameters measured experimentally in references 8 and 20 respectively, and equation (2). Inset shows example spectra at 5 GPa.

Fig. 1 shows the evolution of the G peak of unsupported graphene samples to 8 GPa, with our results for graphite shown for comparison. We observe a slightly larger shift of the graphene G peak to higher wavenumbers at pressure than is the case for graphite. Also in Fig. 1, we calculate the pressure-induced shift of the G peak using the Grüneisen parameter for graphene obtained from the uniaxial strain experiments of Ref. 8 ($\gamma_{E2g}$ = 1.99), and using the Grüneisen parameter for graphite from Ref. 20 ($\gamma_{E2g}$ = 1.59 using the definition of equation (2)). We find good agreement with the experimental data for graphene and graphite respectively. The small difference in behaviour between graphene and graphite observed in Fig. 1 is not necessarily mechanical in origin. It could instead be due to doping of the graphene samples by the nitrogen pressure-transmitting medium at high pressure, an effect which would not be observable for bulk graphite. In any case, the reported pressure-induced shifts of the Raman G peak for graphite vary, ranging from 4.1 cm$^{-1}$GPa$^{-1}$ [25] to 4.7 cm$^{-1}$GPa$^{-1}$ [20].

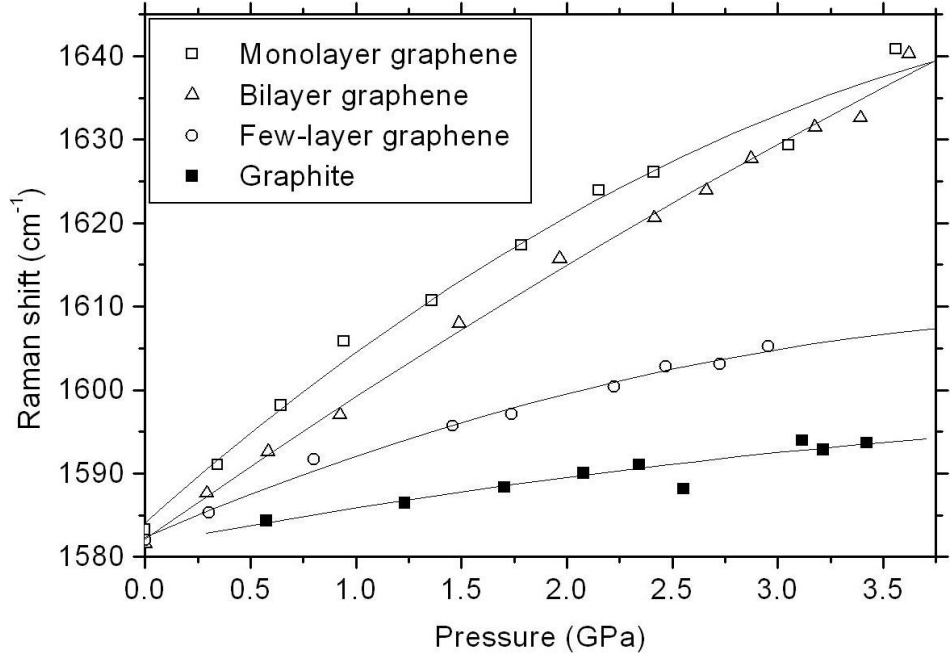

Figure 2. The evolution of the Raman G peak is shown to ≈3.5 GPa for monolayer (open squares), bilayer (open triangles) and few-layer (open circles) graphene on silicon and free-standing graphite (filled squares). Our results for free-standing graphite are virtually identical to those we obtain for graphite on silicon (see Figure 3). Lines are polynomial fits, intended only as guides to the eye.

In Fig. 2 we show our data for the Raman G peak of monolayer, bilayer and few-layer graphene on silicon and freestanding graphite on silicon. A clear trend is observed that the shift of the G peak to higher wavenumber with applied pressure is larger for thinner flakes of graphene. We observe the same trend for the D* peak (see supplementary material). To confirm this trend we loaded two flakes of graphene of different thickness into the diamond anvil cell simultaneously (Fig. 3). The Raman G peak from the flake that was visibly thicker (see photograph in the inset) shifted to higher wavenumbers at a slower rate with applied pressure. Fig. 3 also shows our data for the Raman G peak of a thin layer of graphite on silicon deposited using the same mechanical exfoliation technique as is used for the graphene samples, and for free-standing graphite. The observed pressure-induced Raman shifts for the supported and free-standing graphite are found to be very similar.

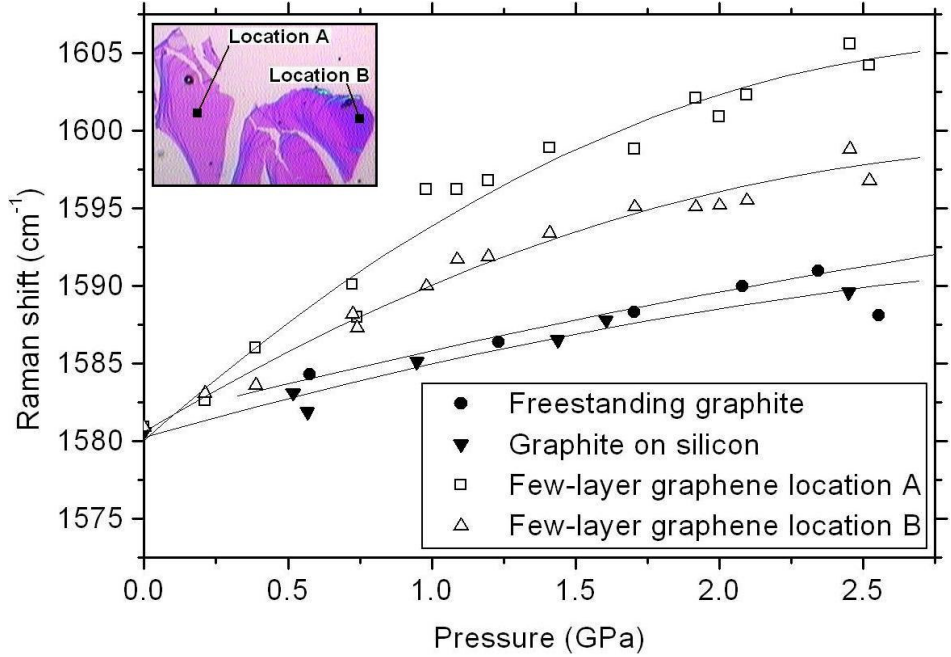

Figure 3. The evolution of the Raman G peak is shown for the few-layer graphene flakes of different thicknesses shown in the inset (the flake at location B was thicker than at location A, shown in colour online). For comparison, data for freestanding graphite and graphite on silicon are presented. Data for graphene and graphite on silicon in this figure were taken with 4:1 methanol-ethanol solution as pressure transmitting medium. Lines are polynomial fits, intended only as guides to the eye.

Our calculations of the pressure-induced shifts of the graphene Raman peaks (equations (1) and (2)) can be extended to the case of graphene on the silicon / $SiO_2$ substrate. In these calculations, we assume that the graphene, $SiO_2$ and silicon remain bonded. The silicon layer is 100 μm thick, the $SiO_2$ layer is 300 nm thick and the graphene layer is < 1 nm thick. We adopt the approach used in the extensive literature on the fabrication and high pressure study of III-V and II-VI semiconductor materials grown epitaxially on a substrate. If the epilayer is sufficiently thin compared to the substrate, it is assumed that the compression of the entire sample is determined by the compressibility of the substrate. See for example Refs. 27 and 28. We therefore assume that the compression of our sample is determined by the compressibility of the silicon substrate. We derive the linear compressibility of silicon at low pressure from the bulk modulus. The bulk modulus $B_0$ is related to the compression of the Si-Si bonds as follows:

$$B_0 = -V \frac{dP}{dV} \tag{4}$$

$$V(P) = V_0 \left(\frac{r(P)}{r_0}\right)^3 \tag{5}$$

The Si-Si bond length at zero pressure is $r_0$ and at arbitrary pressure $r(P)$. From equations (4) and (5) we can derive an approximate relation for the linear compressibility at low pressure and use the bulk modulus from Ref. 26, $B_0 = 97.88$ GPa, to calculate it's value.

$$\frac{d}{dP}\left(\frac{r}{r_0}\right) \approx \frac{1}{3B_0} \approx 0.0034 GPa^{-1} \quad (6)$$

To calculate the pressure-induced shifts of the intra-layer Raman modes of graphene on the substrate we replace the C-C bond lengths a(P) and $a_0$ in equation (2) with those for silicon, r(P) and $r_0$, where r(P) is calculated using the linear compressibility obtained in equation (6).

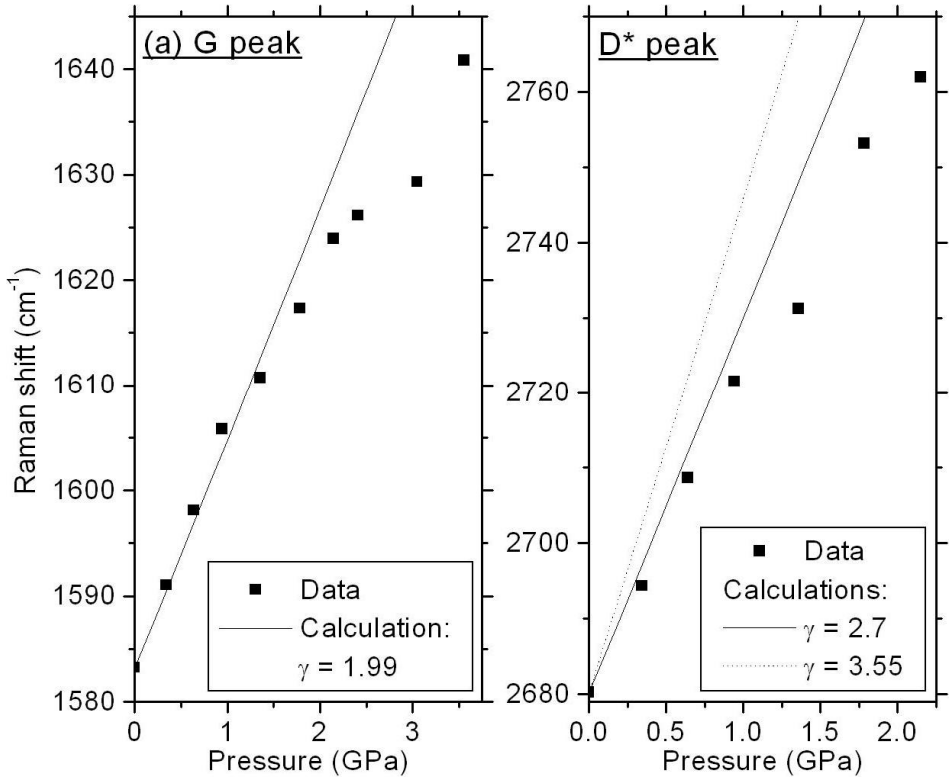

Figure 4. The evolution of the G (a) and D* (b) Raman peaks for monolayer graphene are shown to 3.6 and 2.3 GPa respectively (black squares). In (a), the solid line is the calculated Raman shift for monolayer graphene adhering perfectly to a silicon substrate using the Grüneisen parameter obtained from recent investigations on graphene under uniaxial strain [8] ($\gamma_{E2g} = 1.99$). In (b), the solid line is the calculated Raman shift using the Grüneisen parameter obtained from the density-functional theory calculations for graphene [8] ($\gamma_{A1g} = 2.7$), and the dashed line is the calculated Raman shift using the Grüneisen parameter obtained from the investigations on graphene under uniaxial strain [8] ($\gamma_{A1g} = 3.55$).

In Fig. 4 we compare our data for the G and D* peaks of monolayer graphene on silicon to our calculations using the Grüneisen parameters obtained in recent experiments on graphene under uniaxial strain [8]. At lower pressures, we find good agreement in the case of the G peak. However, in the case of the D* peak our results show better agreement with those on graphite under hydrostatic pressure [21] ($\gamma_{A1g} =$

2.84) and with those of the density-functional theory calculations for graphene [8] ($\gamma_{A1g}$ = 2.7) than with those of the uniaxial strain experiments on graphene [8] ($\gamma_{A1g}$ = 3.55). However, as discussed in ref. [8], this can be explained by the origin of the D* Raman peak in a double resonance Raman process. In an experiment under uniaxial strain, the relative positions of the dirac cones in the graphene electronic bandstructure are changed, so the double resonance condition and actual phonon probed in the Raman measurements will also change. However, in our experiments the in-plane compression is biaxial and the effects due to the relative movement of the dirac cones are absent. This is why our data are in agreement with the hydrostatic pressure experiments on graphite [21] instead of the experiments on graphene under uniaxial strain [8].

At higher pressures, both the G and D* Raman peaks shift to higher wavenumbers with applied pressure at a lower rate than predicted. This could be due to debonding between the different layers of our sample (graphene, $SiO_2$ and silicon). Since the $SiO_2$ and silicon are covalently bonded while the graphene is attached to the $SiO_2$ only by Van der Waal's forces, poor adherence between the graphene and the $SiO_2$ at higher pressures is most likely. Using equation (6) we see that 2 GPa corresponds to a compressive biaxial strain of 0.68% for the graphene on silicon / $SiO_2$. However, in high pressure studies of carbon nanotubes, the molecular organization of the pressure-transmitting medium on the surface of the nanotubes has been shown to have an observable effect on the Raman spectra at high pressure [29]. Our data on few-layer graphene displays a clear trend, that the pressure-induced shifts of the G and D* Raman peaks are larger for thinner samples. This suggests that thicker samples do not adhere as well to the substrate under compressive stress as a monolayer.

It is interesting to note that the low-pressure compressibility of materials could in principle be measured using only optical spectroscopy by depositing a monolayer of graphene on the surface of the material, and using Raman spectroscopy of the monolayer to measure the strain in the material under investigation.

In conclusion, we have studied graphene at high pressure for the first time, in order to confirm key mechanical characteristics of the material. These are needed due to it's potential applications in nanoelectronics, in which the intentional use of strain could improve device characteristics, and the current interest in intrinsically stressed graphene samples, such as graphene grown epitaxially on SiC. Our comparison of supported and unsupported graphene samples demonstrates that the compression of the graphene is initially (i.e. at low pressures) determined by the compressibility of the substrate. The good adherence of monolayer graphene to the Si / $SiO_2$ substrates in our experiments under hydrostatic pressure, and to the polydimethylsiloxane, polyethylene terephthalate and perspex substrates in the recent uniaxial strain experiments [4, 8], demonstrates the potential application of graphene as a strain sensor.

The authors would like to acknowledge the help of Malachy McGowan (School of Electrical and Electronic Engineering, University of Manchester) in the preparation of the graphene samples. We would also like to thank Dr. Chrystéle Sanloup (School of Geosciences, University of Edinburgh) and Prof. David Dunstan (Physics Dept., Queen Mary, University of London) for advice about the calculations.

# Supplementary information

**Preparation of graphene samples**

Supported graphene samples were prepared using the mechanical exfoliation technique [1], on 100 µm thick silicon wafers coated with a 300 nm thick $SiO_2$ layer. The thickness of various graphene flakes was identified with Raman spectroscopy [30]. The silicon wafer around these flakes was cut using a focussed infrared laser, producing ≈ 300 µm diameter discs of silicon with graphene flakes in the middle. Monolayers and bilayers were cut in a nitrogen atmosphere to prevent oxidation of the samples. After checking with Raman spectroscopy that the graphene had not been oxidised during the cutting process [31] the samples were placed in the DAC. The nitrogen pressure-transmitting medium was loaded cryogenically.

Unsupported graphene samples were prepared by liquid-phase exfoliation of graphite [22]. Using the procedure described in Ref. 22, a 0.005 mg/ml dispersion of graphene in NMP surfactant was prepared. The solution was filtered onto alumina membranes with 20 nm pores, resulting in a graphene film approximately 100 nm thick. Due to aggregation during the film formation phase, these films are a mixture of monolayer, bilayer and few-layer graphene and also contain some nanographite. However, the Raman spectrum in Fig. S1 includes a D* band that is characteristic of few-layer graphene rather than graphite [30] so the amount of nanographite present must be small. Samples of this film were removed from the membrane and placed directly in the DAC. The nitrogen pressure-transmitting medium was loaded using a high pressure (1 kbar) gas loader.

**Example Raman spectra**

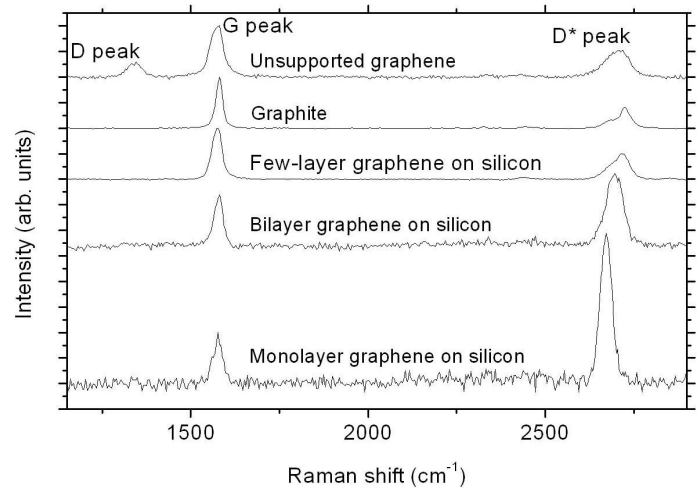

Figure S1. Raman spectra of our samples at ambient pressure. Spectra of monolayer, bilayer and few-layer graphene on silicon are shown after the cutting process described here. The top spectrum is unsupported graphene produced by liquid phase exfoliation [22] after removal from the membrane. Despite this sample containing a small amount of nanocrystalline graphite the D* peak is still that which is characteristic of graphene rather than graphite [30], and the presence of a D peak indicates that the graphene flakes are very small [22].

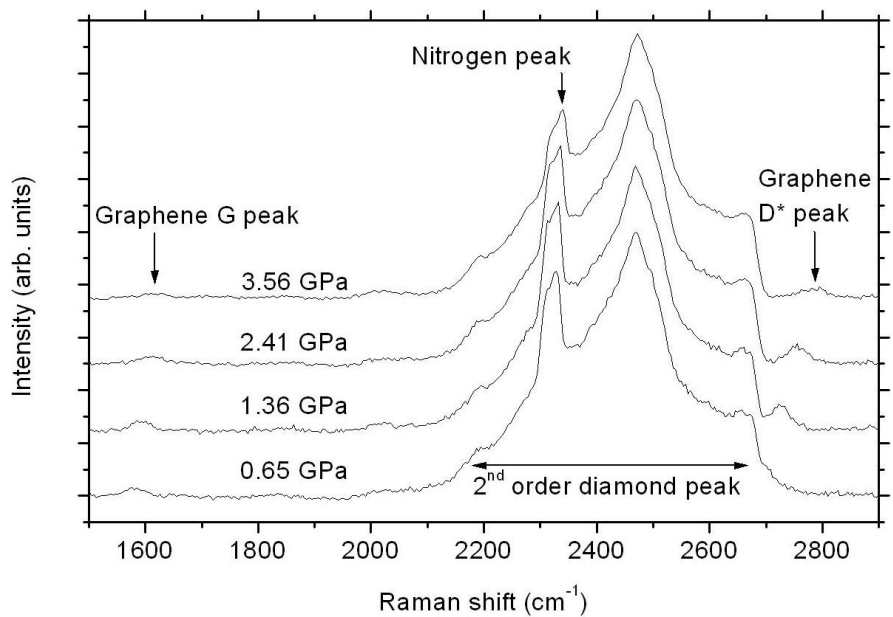

Figure S2. Evolution of the G and D* peaks of monolayer graphene on a silicon / SiO$_2$ substrate with increasing pressure. At atmospheric pressure the D* Raman peak from the graphene overlaps partially with the 2nd order Raman peak from the diamond anvil in the high pressure cell, so a Raman spectrum taken in a part of the cell away from the graphene sample was subtracted from the Raman spectrum of the sample. However, at higher pressures the D* graphene peak separates from the 2nd order diamond Raman peak as it shifts to higher wavenumbers with applied pressure at a much faster rate. The nitrogen Raman peak is present as nitrogen is used as the pressure-transmitting medium.

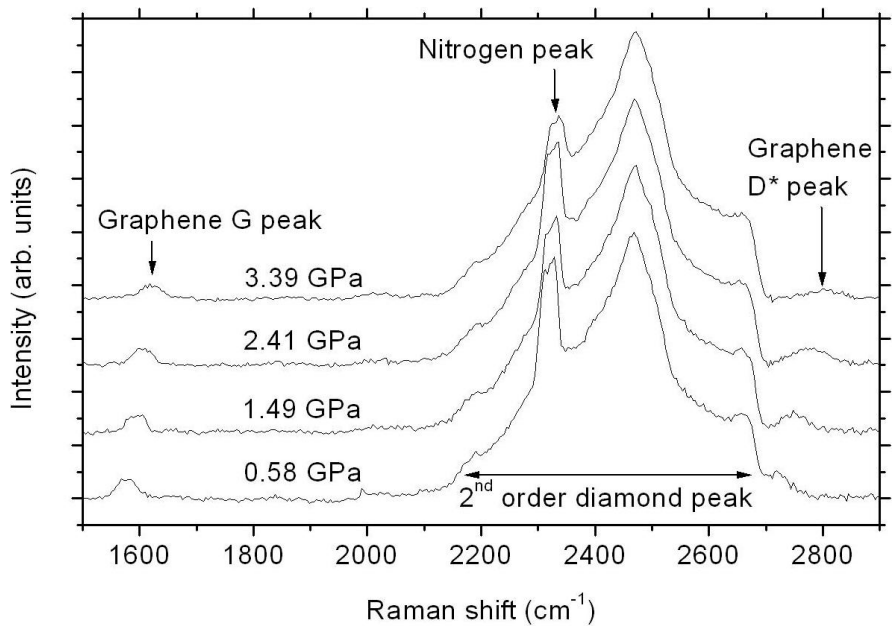

Figure S3. Evolution of the G and D* peaks of bilayer graphene on a silicon / SiO$_2$ substrate with increasing pressure.

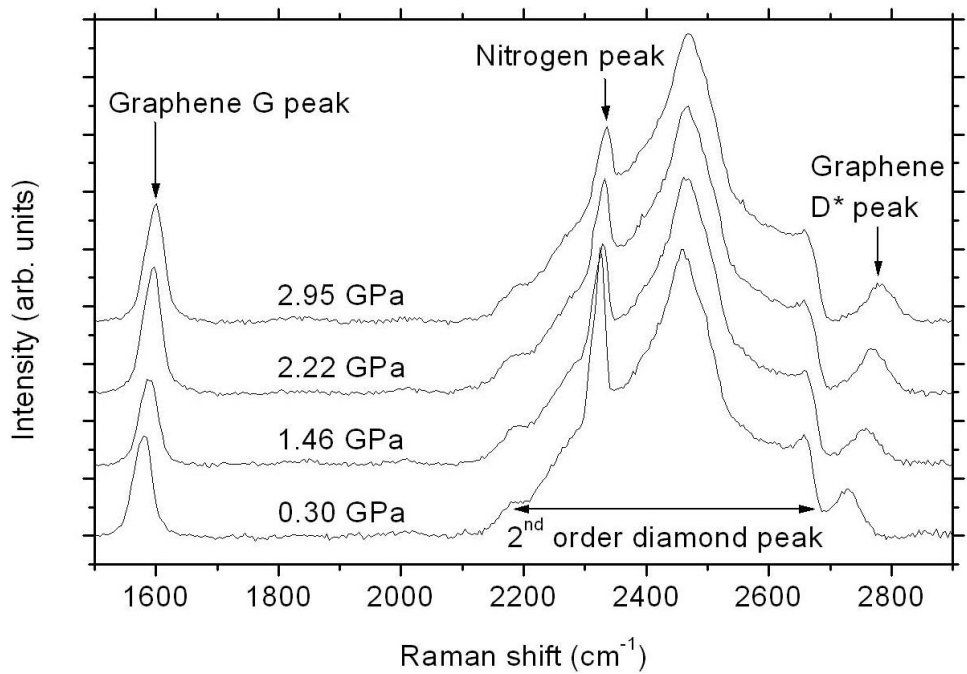

Figure S4. Evolution of the G and D* peaks of few-layer graphene on a silicon / $SiO_2$ substrate with increasing pressure.

**Evolution of D\* Raman peak position at high pressure**

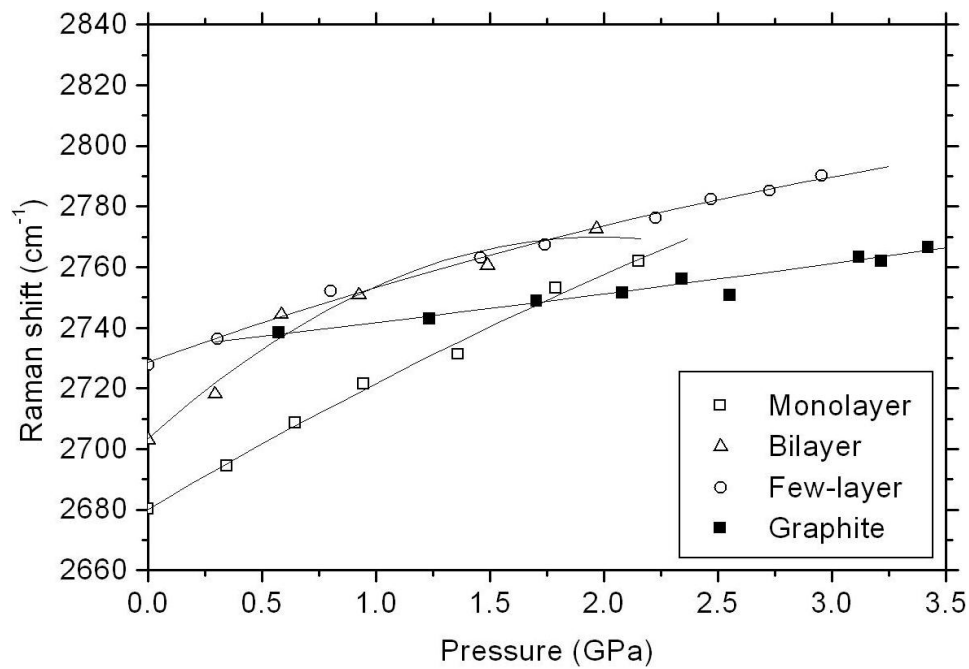

Figure S5. The evolution of the Raman D* peak with increasing pressure is shown for monolayer, bilayer and few-layer graphene on silicon and for free-standing graphite. The same trend is observed as for the G peak in figures 2 and 3 in the main paper - the rate of shift with applied pressure is greater for thinner graphene samples. Lines are polynomial fits, intended only as guides to the eye.